\documentclass[10pt, conference, letterpaper]{IEEEtran}
\usepackage{mathptmx}
\usepackage{graphicx}
\graphicspath{ {figs/} }
\usepackage{blindtext}
\usepackage{mathtools, color}

\usepackage{layout, color}
\newcommand{\fcorp}{Friend Corpus\xspace}
\newcommand{\bulk}{bulk\_extractor\xspace}

\DeclarePairedDelimiter\abs{\lvert}{\rvert}%

\renewcommand{\title}[1]{\noindent\textbf{#1}\bigskip\\}
\renewcommand{\author}[1]{\noindent #1\bigskip\\}

\begin{document}


\title{Creating and understanding email communication networks to aid digital forensic investigations}
\author{Michael McCarrin, Janina Green\footnote{The current research includes excerpts from~\cite{green2016constructing}, which is one of the author's thesis.} and Ralucca Gera \bigskip\\
}

\begin{abstract}
Digital forensic analysts depend on the ability to understand the social networks of the individuals they investigate. We develop a novel method for automatically constructing these networks from collected hard drives. We accomplish this by scanning the raw storage media for email addresses, constructing co-reference networks based on the proximity of email addresses to each other, then selecting connected components that correspond to real communication networks. We validate our analysis against a tagged data-set of networks for which we determined ground truth through interviews with the drive owners. In the resulting social networks, we find that classical measures of centrality and community detection algorithms are effective for identifying important nodes and close associates.   

\end{abstract}

\IEEEpeerreviewmaketitle


\section{Introduction}\label{sec:Introduction}
Email data stored on hard drives provides a wealth of information about the users of the devices on which it is stored. In a digital forensic investigation, email headers and contact lists can often be of greater importance than the content of the email in that they identify a user's associates and provide a lower bound on the frequency with which the user communicated with them. Several straightforward methods of extracting email addresses are available. An analyst willing to take the time to inspect a drive manually can likely locate many obvious sources of email addresses, including address books or addresses stored in email headers. The primary drawback of this technique is scalability: it is only effective if the quantity of drives requiring analysis is relatively small such that analyst time can be devoted to each. Commonly, this is not the case; many organizations require analysis of drives on a scale that far outstrips the capacity of their examiners. Furthermore, the work becomes more challenging if the email is stored in an unusual location or an unknown file format, or resides in unallocated space (as a result of naive deletion or fast-formatting).

A more scalable approach, developed by Garfinkel~\cite{garfinkel2006forensic}, is to scan the drive linearly from the first addressable sector to the end and extract all email addresses in bulk. This process can be automated easily and is robust against variations in file system, format or storage location (though compressed or encoded data must be handled specially). Unfortunately, it tends to produce a high number of false positives---email addresses that are present on the drive due to software installations, documentation, security certificate stores and other sources, but which have no direct correlation to the social network of the device owner. The presence of these false positives again requires analyst time to cull through and identify addresses used for actual communication between the drive owner and his or her associates. Because a given drive can have tens of thousands of such false positives, the need for manual inspection can erase any efficiency gains offered by the extraction method.

We improve this method by constructing co-reference networks based on the byte offset of the email addresses in storage. Our approach relies on the tendency of storage devices and file systems to store related data in the same area for purposes of optimizing performance, a consequence of the locality principle~\cite{denning2005locality}.  Because co-located data tends to be related, addresses that appear near each other can be treated like names appearing in the same document, and the resulting co-reference networks tend strongly to group communication-related email addresses in separate connected components from addresses that are artifacts of installed software. Furthermore, because frequency of email or chat communication corresponds to repetition of co-located pairs of addresses on the drive, components comprised of email addresses used in communication reveal information about the social network of the drive user. This technique preserves the advantages of the linear scan while greatly reducing the manual effort required to arrive at a human-readable report.  

The remainder of this paper proceeds as follows: Section~\ref{sec:relatedWork} gives a summary of work in digital forensics and social network analysis that our research seeks to extend. Section~\ref{sec:constructingnetworks} describes our methodology for building social networks from hard drives and Section~\ref{sec:datasets} discusses the datasets we used to test our approach.  In Section~\ref{sec:results} we characterize some exemplary networks produced from our data. Finally, in Section~\ref{sec:conclusionfuture} we present our conclusion and propose areas for future work.

\section{Related Work}\label{sec:relatedWork}
 
Our work combines methods developed in two distinct areas of research: forensic analysis of email addresses found on secondary storage media, such as hard drives, and network science analysis of co-reference networks created from documents. We begin with a discussion of work in digital forensics.

Early methods of conducting digital forensic analysis focused primarily on the needs of a single examiner working on a case-by-case basis~\cite{garfinkel2010digital}. To address the increasing volume of data requiring analysis, Garfinkel developed an automated scheme for forensic feature extraction and cross-drive analysis~\cite{garfinkel2006forensic}. His method extracts forensic artifacts, such as email addresses, using regular expressions to scan the raw media without reference to the file system, then attempts to correlate related drives by clustering based on the presence of common artifacts. 

Garfinkel proposes social network identification as a potential application of this approach, but offers no methods for constructing such networks. Further, his experiments reveal that many artifacts produced by this method, such as email addresses extracted from common security certificates, introduce considerable noise and must be filtered out before the correlation step. Garfinkel later developed context-sensitive ``stop lists'' to attempt to solve this problem~\cite{garfinkel2013digital}.

Because comprehensive stop lists are large and quickly go out of date, Rowe et al.\ attempt to enhance the effectiveness of Garfinkel's whitelist approach using a Bayesian filter to find other email addresses similar to those on the list. They then use $k$-means clustering to partition drives into groups, but they provide no independent verification that the resulting groups are meaningful~\cite{rowe2016making}.

In the domain of social network analysis, {\"O}zg{\"u}r et al.\ introduces a method for building social networks based on co-occurrence of names in articles in the news media. Their algorithm for constructing these networks regards all names as nodes.  The edge are between any two names that appear in the same article, adding weight to edges of nodes that co-occur multiple times. They show that applying this technique to the Reuters-21578 corpus~\cite{lewis2003reuters} produces social networks of public figures in which leaders and key relationships are identified by topological properties~\cite{ozgur2008co}. Their process for constructing the network relies on a hybrid of manual and automated analysis in which names were identified manually in $10,000$ of the $21,578$ news articles in their corpus.  They extract from the rest using a named entity recognition tool, the output of which is manually reviewed. This hybrid process is analogous to the approach taken by many tool-assisted digital forensic analysts. Namely, they proceed either by manual examination of devices or by performing extensive review of the output of forensic artifact extraction tools such as bulk\_extractor.

Acevedo-Aviles et al.\ make use of a similar method of network construction in their VizLinc tool.  They leverage several publicly available tools, including the Stanford Named-Entity-Recognizer (NER), to create a single automated pipeline from ingestion to analysis and to draw in other relevant data, such as geographic information~\cite{acevedovizlinc}. At a high level, this pipeline approach resembles Garfinkel's proposed process for extracting digital artifacts, but with the addition of an interactive graphical tool that presents a final unified display. The authors test their system against a subset of the New York Times Annotated Corpus~\cite{sandhaus2008new}, but do not perform a quantitative evaluation of its effectiveness. 


In the current work we seek to advance the state of the art in automated forensic analysis by treating proximate areas of hard drive storage as ``documents'' and forensic artifacts (namely email addresses) as ``named entities'' allowing us to incorporate existing techniques for deriving social network information from document collections. 

\section{Constructing Networks}\label{sec:constructingnetworks}
We begin by extracting email addresses from storage media using the \bulk utility. This produces a flat file of email addresses and the byte offsets at which they were found on the drive. We sort this list by byte offset, then construct a network following an algorithm very similar to that used by {\"O}zg{\"u}r et al.:
\begin{enumerate}
    \item Create a new node for each distinct email address.
    \item Define a window size, W.
    \item Place an edge between all email addresses $e_{i}$ and $e_{j}$, where the difference
          between their corresponding offsets is less than W. That is,
          $\abs{o_{i} - o_{j}} < W$.
    \item If an edge exists between the email address pair already, increment its weight by one. 
\end{enumerate}

Note that in the case of supported compression types, \bulk performs a recursive unzip. This allows us to recover many additional email addresses that would otherwise be unrecognizable in compressed form, but requires special handling, since there is not a one-to-one mapping between the offsets of compressed data and the offsets on the drive. \bulk handles this scenario by representing the byte-offset field as a ``forensic path,'' a hyphen-delimited tuple that
begins with the offset of the compressed file itself, followed by the type of compression detected (for example GZIP), followed by the offset into the uncompressed version of the file.

In our implementation we handle these complex forensic paths by splitting into a ``base'' and an ``offset,'' where the offset is the final integer representing the location in the uncompressed file and the base is all preceding parts of the forensic path. For the majority of features that were extracted without decompressing data, the base value is the empty string. The distance between features with identical bases is calculated by subtracting offsets, whereas the distance between features with different bases is treated as infinite (i.e.
an edge is never placed between features with different bases, regardless of the choice of window size). This choice is reasonable because the transition from decompressed data to compressed data on a drive usually denotes a change of context.
 
\section{Data Sets}\label{sec:datasets}
We use two datasets to test our approach: the M57 corpus~\cite{woods2011creating} and a set of drive images we collected specifically for this project, referred to here as the \fcorp. Following Garfinkel et al.~\cite{garfinkel2009bringing}, we evaluate the quality of a test data set for digital forensics research in terms of access to ground truth, verisimilitude, release restrictions and scale. Here ``verisimilitude'' is the quality of mirroring the nature and complexity of forensic data encountered by examiners sufficiently to ensure that experimental results may be generalized to operational circumstances, and corresponds to the distinction Garfinkel et al. make between ``real'' and ``realistic'' data. Unfortunately, no existing data sets meet our research needs with respect to all of these categories at once. We rely primarily on our in-house dataset because knowledge of ground truth and verisimilitude were high priorities; in future work, we hope to extend to a larger corpus.

\subsection{M57 Patents Scenario Drives}\label{sec:m57drives}
The M57 corpus is a collection of drives created as part of the M57 Patent Case Scenario~\cite{woods2011creating}. It is an artificial data set, in that it was created for educational purposes by researchers following a scripted scenario. The scenario was performed by a team of researchers at the Naval Postgraduate School in 2011. As a result, ground truth is well-documented, and the data is publicly releasable\footnote{http://digitalcorpora.org/corpora/scenarios/m57-patents-scenario}. However, despite considerable effort invested in making this data set realistic, we find it lacks the complexity of disk images collected from non-academic users, and the small scale of the data may raise questions as to the generality of conclusions based on it.


\subsection{\fcorp}\label{sec:frienddrives}
The \fcorp is a set of ten real drive images collected from seven volunteers as part of an IRB protocol established for the purposes of this study~\cite{green2016constructing}. The characteristics of the collected images are summarized in Table~\ref{tab:frienddrives}. The devices imaged included personal laptops and workstations running both Windows and OSX. The protocol does not permit release of the data, and the scale of the study is limited, but we can share properties about the data. This data set has the advantage of combining real user data with ground truth acquired by interviewing the owners of the devices post analysis for validation. To assemble this collection, we advertised to staff and students at our university, imaged and analyzed their machines, then created visualizations of resulting networks to present to the device owners.

With the exception of $d5$, which was a small SSD with limited usage, all drives produced disconnected graphs containing a large number of connected components, and some number of singletons. We generally discarded singletons, since we lacked an analysis strategy to incorporate them. Also, due to limited interview time with the device owners, we focused on only the top twenty components with the highest order. However, these included the vast majority of the nodes in the graphs, since the order tended to fall off rapidly. Table~\ref{tab:nodecoverage} lists the percentage of each network covered in the top twenty components.

We manually reviewed the resulting 200 components and divided them into categories depending on whether they appeared to contain relevant social network data, appeared to be irrelevant artifacts of installed software, or were of unclear origin. We labeled these three categories of graphs ``Useful,'' ``Not Useful'' and ``Uncertain,'' respectively. We then conducted interviews with the owners in which we asked them to evaluate and comment on our interpretation of their networks. This approach allowed us to perform a detailed verification of methods developed on the M57 corpus.  Based on our interviews, we identified with high confidence many clusters of email addresses corresponding to associates of the drive owner, and were able to examine properties of these components that aid in the understanding of the relationships between the associates represented by these addresses.

\begin{table}[h]
\begin{center}
\begin{tabular}{ll*{6}{c}}
ID & Operating System & Drive Capacity \\
\hline
d1          & OSX                  & 320G \\
d2          & Windows 10           & 100G \\
d3          & OSX                  & 500G \\
d4          & OSX                  & 320G \\
d5          & Windows 8            & 20G  \\
d6          & Windows 8            & 460G \\
d7          & Windows 7            & 500G \\
d8          & Ubuntu/BackTrack     & 200G \\
d9          & Windows 7 Enterprise & 400G \\
d10         & Windows 7 Enterprise & 150G \\
\end{tabular}
\caption{Drive Capacity \& OS for \fcorp.}
\label{tab:frienddrives}
\end{center}
\end{table}

\begin{table}[h]
\begin{center}
\begin{tabular}{ll*{6}{c}}
ID & Nodes & Edges & \% Nodes Studied & \% Edges Studied\\
\hline
d1          & 55030 & 68594 & 30\% & 77.5\%\\
d2          & 25101 & 30472 & 43.86\% &  89.99\%\\
d3          & 105457 & 172013 & 25.57\% &  43.19\%\\
d4          & 105482 & 107203 & 27.1\% &  69.82\%\\
d5          & 885 & 95 & 8.93\% &  83.16\%\\
d6          & 98086 & 458255 & 31.16\% &  80.21\%\\
d7          & 29727 & 50323 & 45.05\% &  95.03\%\\
d8          & 42452 & 43535 & 35.75\% &  88.39\%\\
d9          & 2623 & 969 & 23.60\% &  95.15\%\\
d10         & 52327 & 74952 & 41.28\% &  88.24\%\\
\end{tabular}
\caption{Percentage of nodes and edges in top 20 connected components of \fcorp}
\label{tab:nodecoverage}
\end{center}
\end{table}

\section{Results}\label{sec:results}

\begin{figure}[!h]
    \begin{center}
    \includegraphics[width = .45\textwidth{}]{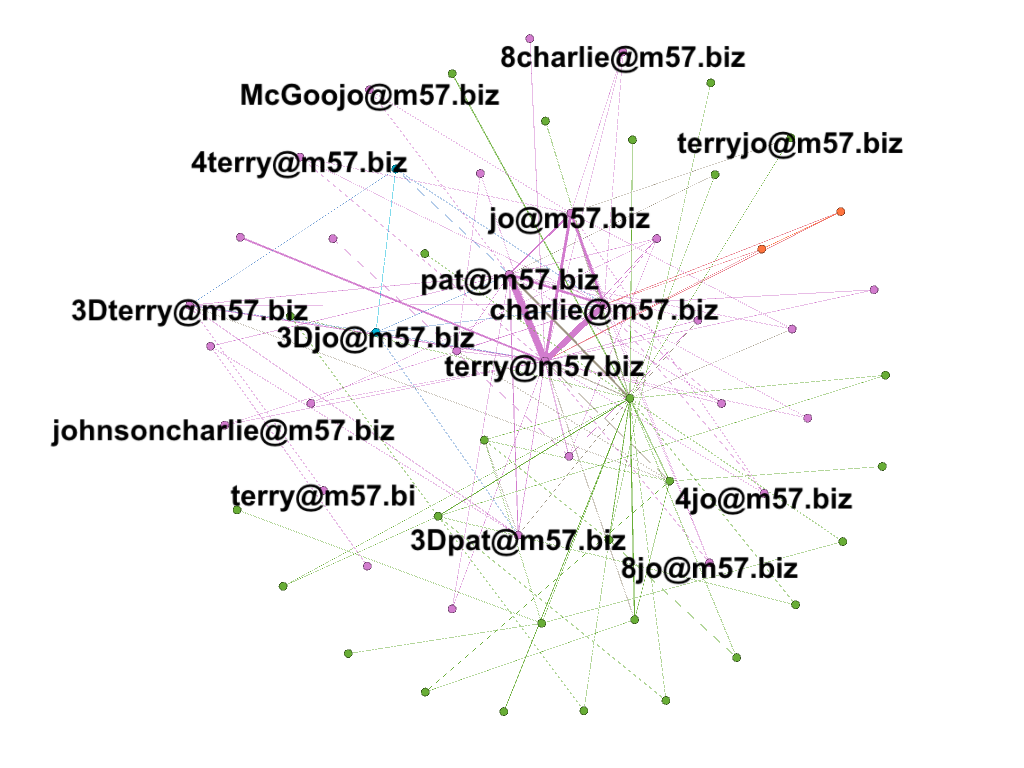}
    \end{center}
    \caption{A large component from a drive in the M57 corpus. Note the heavy weighted lines, which highlight communication between key players in the scenario.}
    \label{fig:m57}
\end{figure}

\begin{figure}[!htbp]
    \centering
            \includegraphics[width=0.45\textwidth]{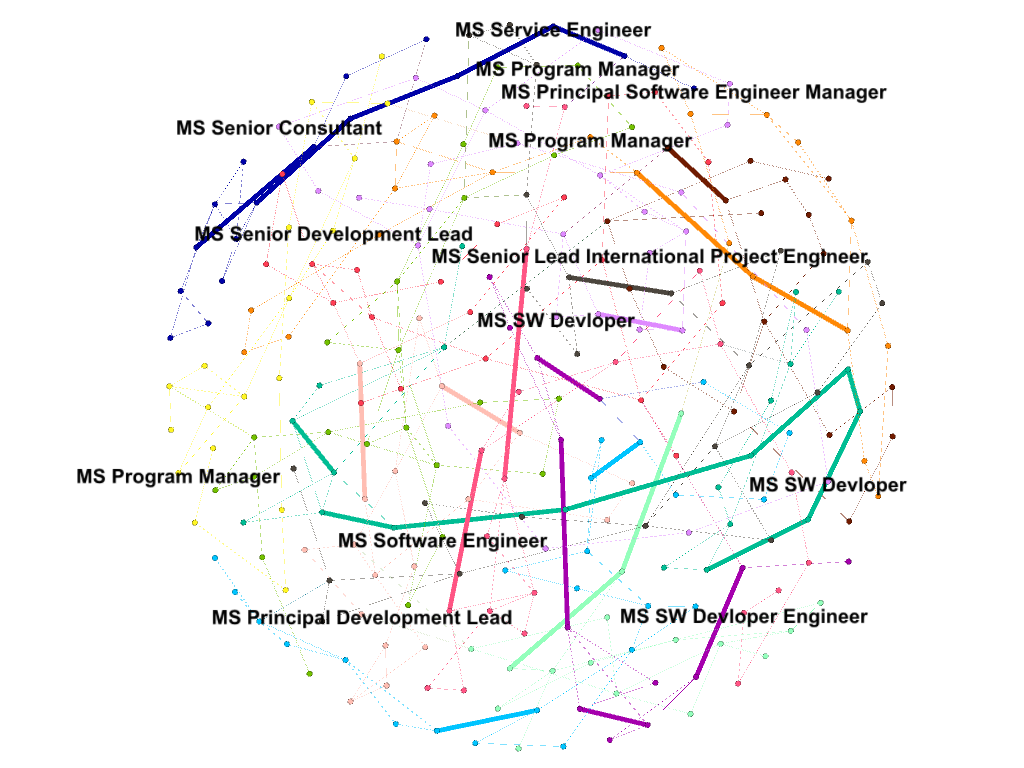}
    \caption[Demonstration of a Component which Displays Microsoft Developers]
            {Component of Microsoft Developers: This component demonstrates a network that was placed in the "Not Useful" bin, but could be used in future works to conduct fingerprinting.  It was found on drives 2, 6, 9, and 10 as the third and/or fourth component. We were able to label the developers job descriptions by conducting web searches. }
    \label{fig:microsoft}
\end{figure}

It was clear from our interviews that the graph construction process effectively separates email addresses involved in user communication from email addresses present on the drive for other reasons. We found no instances of ``mixed'' connected components containing both Useful and Not Useful email addresses. Further, Useful and Not Useful components could often be quickly visually identified. As an example, compare a Useful graph extracted from the $M57$ corpus, shown in Figure~\ref{fig:m57}, with a Not Useful graph of Microsoft developer email addresses shown in Figure~\ref{fig:microsoft}. The former exhibits clear hierarchical structure, and a obvious division between an inner core and a periphery. This structure illuminates the key players in the scenario: the CEO Pat McGoo, IT administrator Terry Johnson, and two patent researchers Jo Smith and Charlie Brown~\cite{m57}. The Microsoft graph, in contrast, is relatively uniform, with no discernible trend behind the heavy-weighted lines.

\begin{figure}[!htbp]
    \centering
            \includegraphics[width=0.49\textwidth]{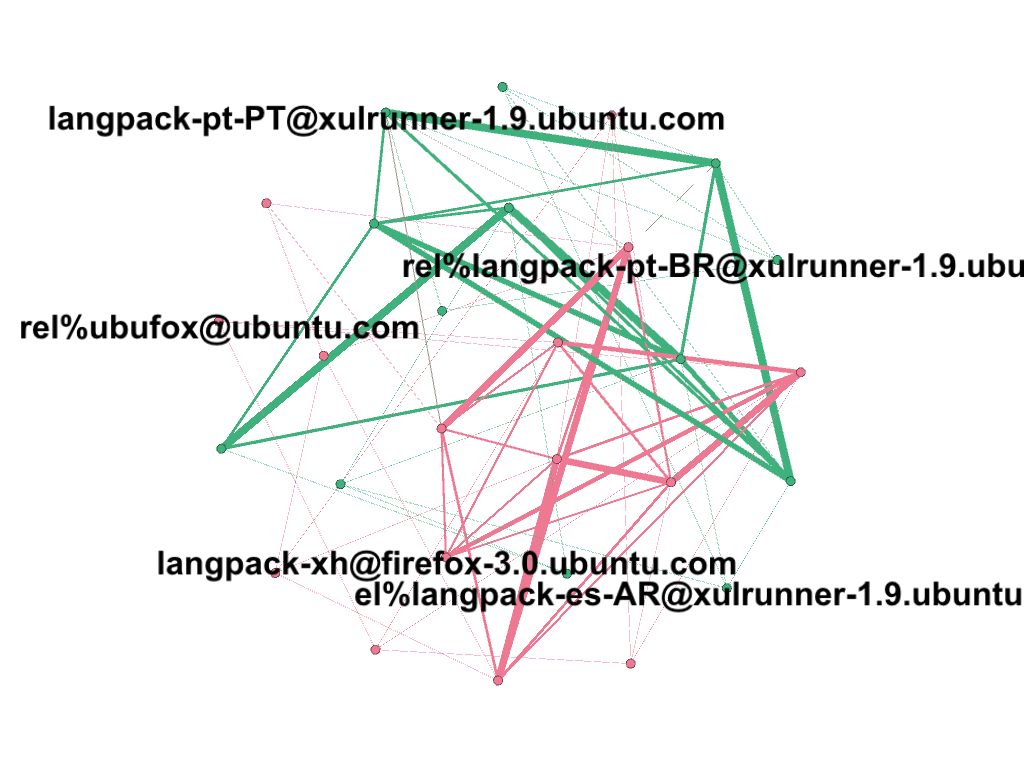}
    \caption{Component created from a USB memory stick that contains a bootable copy of Ubuntu 8.10 Linux~\cite{corpus}. Although we labeled this component ``Not Useful,'' its obvious correlation with the Ubuntu operating system suggests it could be used to fingerprint the installed OS.}
    \label{fig:ubuntu}
\end{figure}

Although this observation holds for many Not Useful components, there were some exceptions. In particular, components created from Ubuntu software package repositories tended to demonstrate a hierarchical structure very similar to communication networks. Figure~\ref{fig:ubuntu} shows one such network. Inspection of the media reveals that the primary source of these email addresses is the ``authors'' line in documentation for programs packaged in the Ubuntu repository. Packages with multiple authors list their names and email addresses in the headers of the text file documentation. Therefore we hypothesize that properties of the corresponding networks resemble true social networks because they are, in fact, co-authorship networks, though this fact is sometimes obscured by language pack extension names that appear with extremely high frequency and resemble email-address structure. Further work is needed to understand how best to automatically differentiate these graphs from Useful components; however, they are sufficiently rare that recognizing the ``ubuntu'' domain might be sufficient as a practical solution.

\begin{figure}[!h]
    \begin{center}
    \includegraphics[width = .45\textwidth{}]{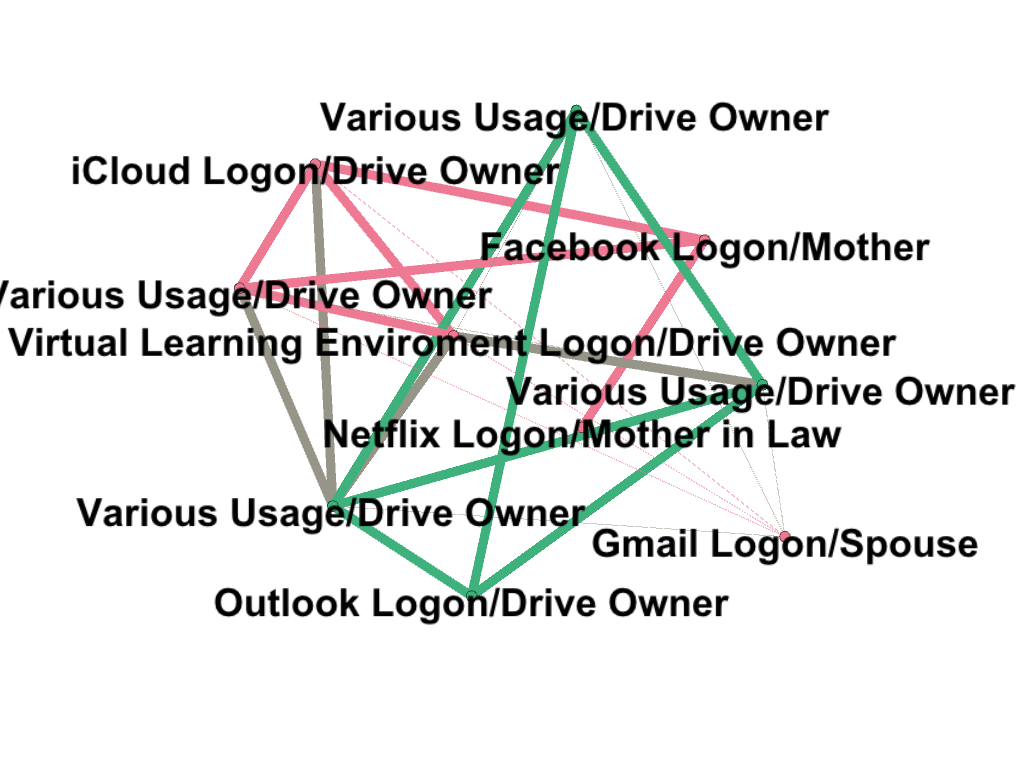}
    \end{center}
    \caption{The network of logon aliases for one user's drive}
    \label{fig:logon}
\end{figure}

One other important edge case in which possession of social-network properties did not exactly correspond to the Useful and Not Useful categories occurred in what we termed ``logon'' networks. Figure~\ref{fig:logon} shows an example of this type of component. The origin of these networks was difficult to decipher due to absence of relevant metadata on the drives, but they appeared to be generated from web cache files. In addition, they differed from other components in that most or all of the nodes corresponded to email addresses that represented the same physical person---usually the drive's primary user. These networks tended to be small, fully collected or nearly fully connected, and to exhibit many heavily weighted edges. Clearly, though they do not provide the kind of rich relationship information contained in a social network graph, they do have considerable value in a forensic investigation, so we have provisionally labeled them as Useful. In the future, their existence may justify the inclusion of an additional category label.

\begin{figure}[!h]
    \begin{center}
    \includegraphics[width = .45\textwidth{}]{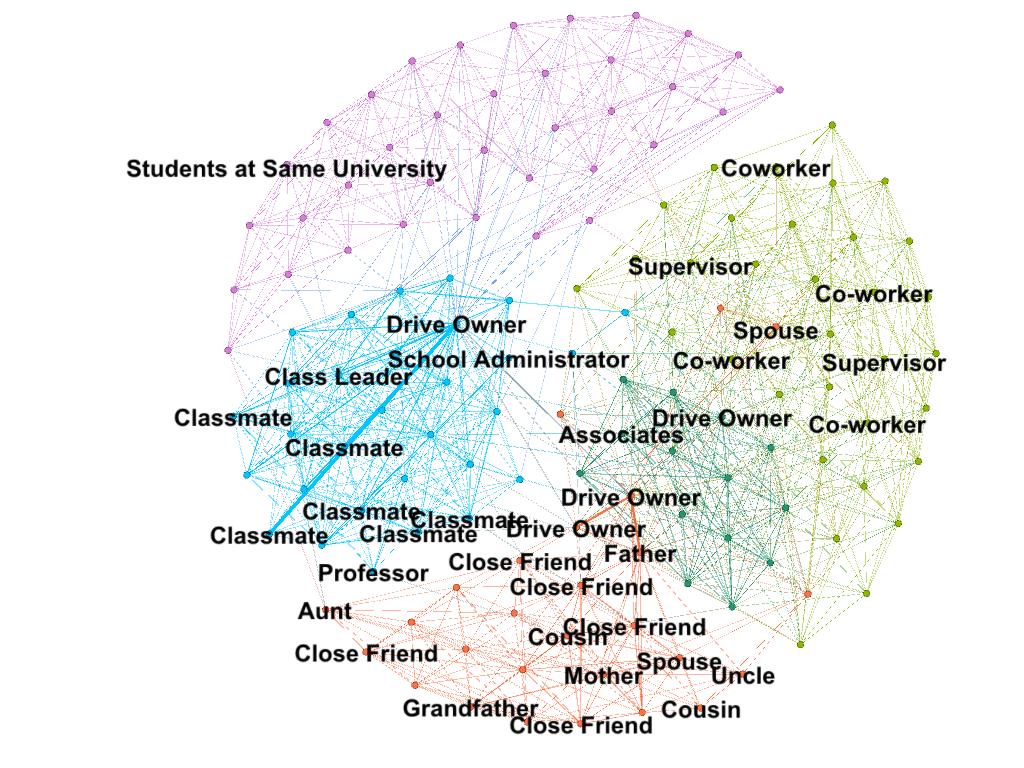}
    \end{center}
    \caption{Social network constructed from a drive in the \fcorp}
    \label{fig:family}
\end{figure}

We were at first skeptical that the outcome of our analysis on the $M57$ drives would remain equally effective when run against real data, but all work so far has confirmed the validity of our approach even in the face of increased complexity. Figure~\ref{fig:family} provides a particularly clear example. In addition, note that community detection naturally and correctly partitions the individuals social groups in this graph.

\begin{figure}[!h]
    \begin{center}
    \includegraphics[width = .45\textwidth{}]{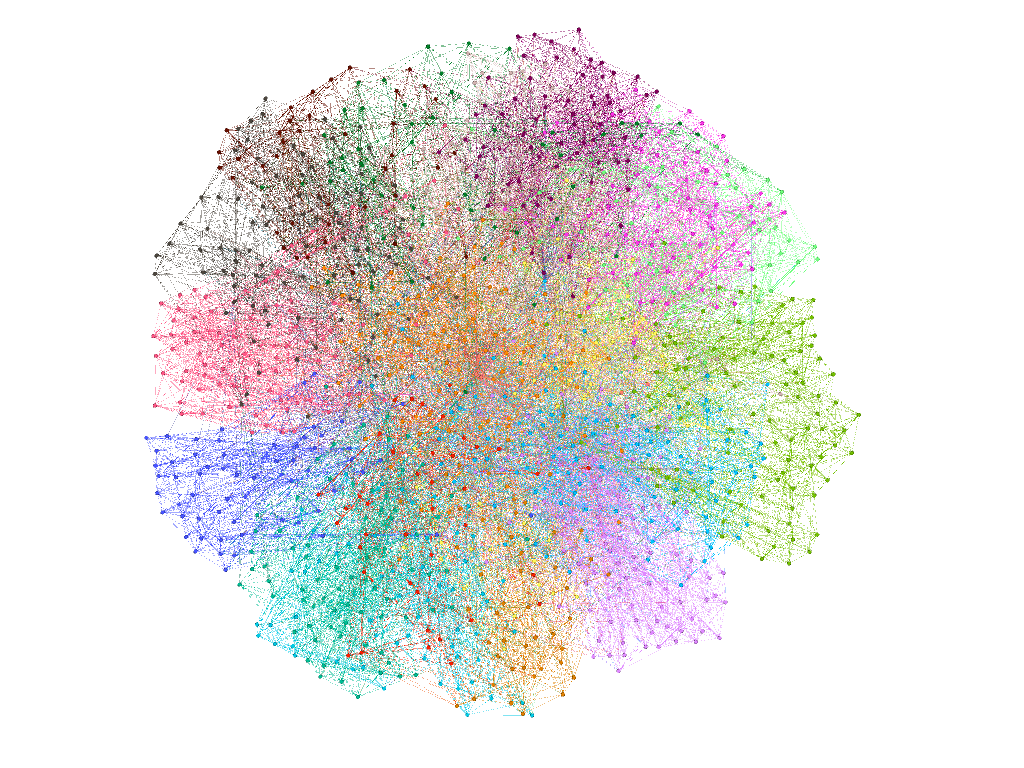}
    \end{center}
    \caption{The drive network of a user who was the administrator for a web-server, colored by community.}
    \label{fig:business}
\end{figure}

A yet more dramatic example of the effectiveness of community detection methods can be seen in Figure~\ref{fig:business}. This storage device belonged to a user who was also the administrator for a web-server running on the device.  Therefore, the owner had email addresses and connections that belonged to personnel who did not access his machine directly.  The server was set up to send copies of emails to the administrator's in-box, when that specific email account had turned on its ``away messages.''  The business was using Dovecot to run their email server and SquirrelMail to access their in-boxes.  (Dovecot is an open source POP3 email server for Linux systems.) Here we see a large and complex component representing email addresses from many different users. However, community detection divides this network into a number of partitions equal to the number of user accounts on the system.

Classic measures of centrality were also effective in identifying important nodes and relationship in extracted Useful graphs. Table~\ref{tab:central} shows a relatively close correspondence between three centrality measures, all of which also agree with ground truth obtained by interviewing the drive owner.

\begin{table}[h]
\begin{center}
\scalebox{0.7}{\begin{tabular}{lll*{3}{c}}
Ranked by Eigenvector       &Ranked by Betweenness & Ranked by Closeness\\
\hline
Drive Owner                 &Drive Owner                 &Drive Owner\\
Father                      &Father                      &Father\\
Classmate A                 &Drive Owner Alt Email 2     &Drive Owner Alt Email 2\\
Classmate B                 &Unknown Email Address 1     &Drive Owner Alt Email 3\\
Drive Owner Alt Email 1     &Classmate Q                 &Coworker different depts\\
Classmate C                 &Classmate K                 &Classmate A\\
Classmate D                 &Department Store            &Drive Owner Alt Email 1\\
Classmate E                 &Classmate A                 &Unknown Email Address 2\\
Classmate F                 &Unknown Email Address 2     &Schoolmate not in same class\\
Classmate G                 &Classmate D                 &Classmate J\\
Classmate H                 &Aunt                        &Classmate Q\\
Classmate I                 &Drive Owner Alt Email 3     &Classmate B\\
Classmate J                 &Classmate L                 &Classmate P\\
Classmate K                 &Cousin                      &Classmate D\\
Classmate L                 &Classmate P                 &Unknown Email Address 1\\
Drive Owner Alt Email 2     &Coworker same dept 1        &Department Store\\
Classmate M                 &Classmate B                 &Spouse\\
Classmate N                 &Classmate M                 &Classmate E\\
Classmate O                 &Classmate E                 &Coworker same dept \\
Classmate P                 &Unknown Email Address 3     &Unknown Email Address 3\\
\end{tabular}}
\caption[Sample Centrality Rankings]{Variations and similarities between rankings of most important nodes by three centrality metrics.}
\label{tab:central}
\end{center}
\end{table}

\begin{table*}[t]
\begin{center}
\begin{tabular}{l*{9}{c}}
Component    & & &Average &Avg. Weighted& & & &Avg. Clustering& Avg. Path\\
ID     &Nodes&Edges&Degree&Degree&Diameter&Density&Modularity&Coefficient&Length\\
\hline
d1c2            &50   &650     &26.00   &166.52      &2     &0.53\%       &.496\%    &.59\%     &1.47\\
d2c2            &10   &43      &8.60    &23.00       &2     &0.96\%       &.08\%     &.95\%     &1.04\\
d3c1            &37   &532     &28.76   &2189.78     &2     &.80\%        &.05\%     &.83\%     &1.20\\
d3c17           &39   &3       &3.13    &3.13        &24    &8.20\%       &69.90 \%  &54.30\%   &8.94\\
d4c2            &20   &159     &15.9    &254.30      &2     &0.001\%      &.21\%     &.84\%     &1.16\\
d4c4            &77   &267     &6.94    &19.80       &6     &.091\%       &.643\%    &.44\%     &3.03\\
d4c7            &123  &2       &2.72    &5.43        &37    &2.20\%       &84.20\%   &57.70\%   &12.87\\
d5c1            &10   &4       &4.80    &8.00        &4     &53.30\%      &17.20\%   &72.20\%   &1.69\\
d5c5            &6    &1       &4.67    &3.00        &2     &33.30\%      &0.00\%    &0.00\%    &1.67\\
d5c6            &5    &1       &1.60    &6.80        &2     &0.40\%       &0.00\%    &0.00\%    &1.60\\
d5c7            &4    &1       &1.50    &1.50        &2     &0.50\%       &0.00\%    &0.00\%    &1.50\\
d5c9            &3    &1       &1.33    &2.00        &2     &0.67\%       &0.00\%    &0.00\%    &1.33\\
d5c14           &3    &2       &2.00    &4.00        &1     &100\%        &0.00\%    &100\%     &1.00\\
d5c17           &2    &1       &1.00    &4.00        &1     &100\%        &0.00\%    &N/A       &1.00\\
d5c20           &2    &1       &1.00    &4.00        &1     &100\%        &0.00\%    &N/A       &1.00\\
d7c1            &6    &15      &5       &14.33       &1     &1\%          &0\%       &1\%       &1\\
d7c2            &9    &3       &4.67    &4.67        &3     &58.30\%      &20.40\%   &76.30\%   &1.50\\
d8c2            &56   &205     &7.321   &87.429      &7     &.13\%        &.484\%    &.70\%     &3.49\\
d9c1            &1240 &13353   &21.537  &412.839     &5     &.02\%        &.82\%     &39\%      &2.96\\
d9c4            &54   &149     &5.52    &11.19       &10    &.10\%        &.69\%     &.65\%     &3.57\\
d9c11           &30   &3       &2.55    &12.36       &11    &12.10\%      &47.40\%   &22.40\%   &4.60\\
d10c1           &33   &447     &27.09   &3215.21     &2     &.85\%        &.25\%     &.85\%     &1.15\\
\end{tabular}
\caption{Metrics Collected from Social Networks Discovered on Disk Images after Filtering to the Network Core, for all components containing more than $200$ nodes.}
\label{tab:core filt stats}
\end{center}
\end{table*}


A summary of metrics that we collected on major networks components is detailed in Table~\ref{tab:core filt stats}. For each, we record node count, edge count, modularity, average weight, eccentricity, and closeness scores, as well as which type of centrality best displayed the drive owner's closer associates (the two metrics tested were Eigenvector and Betweenness centrality). Our early findings suggest that future research into using these metrics to automatically classify components is promising. 
 
\section{Conclusions and Further Directions}\label{sec:conclusionfuture}

We developed a novel method of creating networks to analyze forensic artifacts using network science tools, having the potential to reduce analyst workload. In addition, we perform a preliminary analysis of the properties of our two types of components (those that are not related to social network information and those that are). We used network science tools to screen the graphs, and then to differentiate between the nodes.  Our interpretation of the metrics on the resulting graphs were then validated with the drive owners.

In support of the above work, we also created a test set of drives for which ground truth was determined by interviews with the owners. We tagged these networks based on whether or not the networks demonstrated social networks or if the results were unclear. Networks that were tagged as not containing social network information can be used by those interested in future work to identify and remove networks that do not display social networks. Additionally, networks tagged as interesting can be used by those who would like to prototype algorithms for identifying social network networks or identifying networks that do not display properties of social network networks.

\bibliographystyle{plain}
\bibliography{netscibib}
\end{document}